\documentstyle[prl,aps,epsf,floats]{revtex}
\begin{document}
\draft
\twocolumn[\hsize\textwidth\columnwidth\hsize\csname @twocolumnfalse\endcsname
\title{New interpretation of slave boson mean-field theory  
       of the $t-J$ model:  \\ 
       short-range antiferromagnetic and $d$-wave pairing correlations}
\author{Bumsoo Kyung}
\address{D\'{e}partement de physique and Centre de recherche 
sur les propri\'{e}t\'{e}s \'{e}lectroniques \\
de mat\'{e}riaux avanc\'{e}s. 
Universit\'{e} de Sherbrooke, Sherbrooke, Qu\'{e}bec, Canada J1K 2R1}
\date{August 25, 2000}
\maketitle
\begin{abstract}

   The $t-J$ Hamiltonian is studied in a mean-field approximation 
by taking into account antiferromagnetic and $d$-wave pairing correlations.
Considering the presence of antiferromagnetic fluctuations, the 
weaknesses of a mean-field approximation and the limitation of  
the $t-J$ model near half-filling,
we give a new interpretation to the 
slave boson mean-field theory of the $t-J$ model.
We argue that due to phase coherence-breaking antiferromagnetic
fluctuations and quantum fluctuations,   
superconducting long-range order does not  
appear strictly in two dimensions.
$T_{c}$ resulting from interlayer pairing hopping
can lead to a universal relation,
when $T_{c}$ is
scaled by $T^{\mbox{max}}_{c}$.
Systematic reduction of superfluid density and increase of   
$(\Delta_{d})_{\mbox{max}}/K_{B}T_{c}$ ratio below and near optimal doping 
have their natural explanation in our picture.
A crossover temperature $T^{0}$ found in 
some of magnetic experiments such as NMR
is also easily understood in the present framework.
\end{abstract}
\vskip2pc]
\narrowtext
\section{Introduction}
\label{section1}

     Since the discovery of high temperature superconductivity
in Ln$_{2-x}$Ba$_{x}$CuO$_{4}$ by Bednorz and M\"{u}ller,\cite{Bednorz:1986}
many anomalous features found in these materials have 
attracted considerable attention from condensed matter physicists.
Not to mention the quantitative understanding of various puzzling 
experiments,
even the qualitative understanding of {\it overall picture} of cuprate 
superconductors has been a challenge.
Let us start by discussing a generic phase diagram
of a hole-doped cuprate Ln$_{2-x}$Sr$_{x}$CuO$_{4}$ in the
doping ($x=1-n$) and temperature ($T$) plane.\cite{Almasan:1991}
Near half-filling and at low temperature, antiferromagnetic (AF) long-range
order appears with
$T_{N}=$ 250-300 K at $x=0$.
It is destroyed by $2 \%$ doping concentration.
When $x$ reaches $0.06$, superconducting (SC) long-range order starts to
appear, and it is also destroyed by $30 \%$ doping.
In between them, $T_{c}$ reaches a maximum value of 40 K at $x \simeq 0.16$.
It appears that for most hole-doped cuprates, 
a universal relation\cite{Zhang:1993}
$T_{c}/T^{\mbox{max}}_{c} = 1-82.6(x-0.16)^{2}$ is satisfied.
The SC gap was found to have mainly
$d$-wave character with possibility
of a small mixture of other angular momentum
states,\cite{Levi:1993,%
Shen:1993,Marshall:1996}
in contrast to conventional BCS superconductors\cite{Bardeen:1957}
with an isotropic
$s$-wave gap.

   Various recent experiments also show the existence of a crossover
temperature $T^{*}$ larger than $T_{c}$ in a doping range of
$x=0$ to $x \simeq 0.18-0.19 $.%
\cite{Ding:1996,Loeser:1996,%
Loram:1993,%
Renner:1998,%
Takigawa:1991,%
Homes:1993}
Below this pseudogap temperature $T^{*}$,
the low frequency spectral weight begins to
be strongly suppressed.
Surprisingly the doping dependences of $T^{*}$ and $T_{c}$ are
completely different\cite{Tallon:2000} in spite of
their close relationship suggested by angle resolved photoemission
(ARPES)\cite{Ding:1996,Loeser:1996},
tunneling\cite{Renner:1998} and NMR experiments.\cite{Takigawa:1991}
At optimal doping where $T_{c}$ is maximum, various non-Fermi
liquid (NFL) properties are observed in the normal state.
These include the linear temperature dependence of ab-plane resistivity,
the quadratic $T$ dependence of Hall angle and so on
up to 1000 K.
Far beyond optimal doping, the normal state properties are well
described by the conventional Landau Fermi liquid.
Near optimal doping and underdoping,
superfluid density $n_{s}$ is
systematically suppressed with decreasing doping in spite of
increasing SC gap amplitude.
The resulting $(\Delta_{d})_{\mbox{max}}/K_{B}T_{c}$ ratio is strongly 
violated from the universal BCS value. 
In the overdoped regime, however, the SC properties appear to be well explained
by the conventional weak coupling BCS theory.

    Right after the discovery of high temperature superconductors, 
Anderson\cite{Anderson:1987} first proposed 
the one-band Hubbard model
as the simplest Hamiltonian which might capture the correct
low energy physics of copper oxides.
In that seminal paper, he conjectured that the ground state 
of cuprates at half-filling and presumably away from half-filling as well
may be described by a resonating 
valence bond (RVB) state.
Subsequently Anderson and his co-workers applied a mean-field 
approximation\cite{Baskaran:1987} to the strong coupling limit 
of the Hubbard model, namely,
the $t-J$ model.
The $t-J$ Hamiltonian was already known to be the large $U$ limit
of the Hubbard Hamiltonian under certain assumptions.
The phase diagram obtained by these authors  
and by others\cite{Ruckenstein:1987,Kotliar:1988,Affleck:1988,Dombre:1989,%
Lederer:1989,Ubbens:1992}
has been a starting point for further development 
of the theory such as $1/N$ expansion theory\cite{Grilli:1990}
and gauge theory\cite{Baskaran:1988,Ubbens:1994}
of the $t-J$ model.

   In order to more conveniently handle the no-double-occupancy constraint 
imposed by the $t-J$ model, a slave boson representation of an original  
electron is often used.
In this representation,
an electron is decomposed 
into a spinon (fermion) and a holon (boson),
$c^{+}_{i,\sigma} = f^{+}_{i,\sigma}b_{i}$. 
In slave boson theory of the $t-J$ model,  
typically two mean-field order parameters are considered 
\begin{eqnarray}
\chi_{ij} &=& \langle f^{+}_{i,\sigma}f_{j,\sigma} \rangle  \; ,
                                             \nonumber  \\
\Delta_{ij} &=& \langle f_{j,\uparrow}    f_{i,\downarrow}
                    -f_{j,\downarrow}  f_{i,\uparrow} \rangle
                                                      \; ,
                                                         \label{eq0.1}
\end{eqnarray}
together with $\langle b_{i} \rangle$.
Depending on the vanishing or nonvanishing of $\Delta_{ij}$ and
$\langle b_{i} \rangle$, the doping and temperature plane is divided
into four regions.\cite{Ubbens:1994}
Region I with $\Delta_{ij} = 0$ and $\langle b_{i} \rangle \ne 0$ is 
a Fermi liquid phase. 
Region II with $\Delta_{ij} \ne 0$ and $\langle b_{i} \rangle = 0$ is 
the spingap phase, in which a $d$-wave gap appears in the fermion  
spectrum without Bose condensation of holons.
Region III with $\Delta_{ij} \ne 0$ and $\langle b_{i} \rangle  \ne 0$ 
indicates SC long-range order in physical electrons.
Region IV with $\Delta_{ij} = 0$ and $\langle b_{i} \rangle  = 0$ 
is designated as the strange metal phase, because it shows  
various non-Fermi liquid features.

   In many respects, the slave boson mean-field 
theory of the $t-J$ model\cite{Baskaran:1987,Ruckenstein:1987,Kotliar:1988,%
Affleck:1988,Dombre:1989,Lederer:1989,Ubbens:1992}
has shed some important insight into the microscopic understanding 
of the cuprate superconductors.
This is because 
the predicted phase diagram is, at least, qualitatively
consistent with experiments, and 
the pseudogap is closely related to a spingap,
and furthermore
it starts from the microscopic model 
as opposed to other
phenomenological models.
However, there are also some serious problems 
with the slave boson mean-field theory, 
as noted by Ubbens and Lee.\cite{Ubbens:1994}
One of them 
is that the temperature scale for Bose condensation of holons 
is too high. Furthermore the maximum $T_{c}$, which is determined  
by the two lines 
$\Delta_{ij} \ne 0$ and $\langle b_{i} \rangle  \ne 0$, occurs
at too small doping concentration ($x < 0.06$).
At this doping level,
the SC long-range order even does not appear in cuprate superconductors.
Close to half-filling, several exotic phases have been reported to 
be stable such as mixed phases\cite{Kotliar:1988}
(equivalently $\pi$-flux phases\cite{Affleck:1988}), 
dimerized phases,\cite{Affleck:1988,Dombre:1989} and 
staggered flux phases.\cite{Lederer:1989,Ubbens:1992}
It is unclear whether these states are realized or not 
in cuprates.
In this paper we argue that these problems can be naturally 
resolved, when AF correlations, the weaknesses of a mean-field approximation
and the limitation of the $t-J$ model near half-filling
are properly taken into account.
In addition to that, we show that the universal relation of 
$T_{c}/T^{\mbox{max}}_{c}$, 
systematic reduction of superfluid density and increase of   
$(\Delta_{d})_{\mbox{max}}/K_{B}T_{c}$ ratio below and near optimal doping 
can be naturally explained in our picture.
\section{Formulation}
\label{section2}

   The $t-J$ model is described by the Hamiltonian
where $c_{i,\sigma}$ destroys an electron at site $i$ with spin $\sigma$
on a two-dimensional square lattice
\begin{eqnarray}
 H = &-t& \sum_{\langle i,j \rangle, \sigma} 
       \bigl((1-n_{i,-\sigma})c^{+}_{i,\sigma}
         c    _{j,\sigma}(1-n_{j,-\sigma}) +  \mbox{H.c.} \bigr)
                                             \nonumber  \\
     &+J& \sum_{\langle i,j \rangle}\bigl( 
        \vec{S}_{i} \cdot \vec{S}_{j}-\frac{1}{4}n_{i}n_{j} \bigr)
     -\mu_{0}\sum_{i, \sigma}c^{+}_{i,\sigma}c_{i,\sigma}   \; .
                                                           \label{eq1}
\end{eqnarray}
$t$ is a hopping parameter between nearest neighbors $<i,j>$ and
$J$ denotes superexchange coupling.
Double occupancy of two electrons at the same lattice site is 
forbidden by a projection operator $(1-n_{i,-\sigma})$ in the 
hopping term.
$\mu_{0}$ is the chemical potential controlling the electron density $n$.
$\vec{S}_{i}$ and $n_{i}$ are spin and charge density operators, 
respectively, and they are defined as  
\begin{eqnarray}
    \vec{S}_{i} &=& \frac{1}{2}\sum_{\alpha,\beta}
                 c^{+}_{i,\alpha}\vec{\sigma}_{\alpha,\beta}
                 c    _{i,\beta}  \; ,
                                             \nonumber  \\
    n_{i} &=& \sum_{\sigma}c^{+}_{i,\sigma}c_{i,\sigma}  \; ,
                                                           \label{eq2}
\end{eqnarray}
where $\vec{\sigma}$ is a $2 \times 2$ Pauli spin matrix.
Through decomposition of an electron 
into a spinon (fermion) and a holon (boson),
$c^{+}_{i,\sigma} = f^{+}_{i,\sigma}b_{i}$, 
the $t-J$ model becomes
\begin{eqnarray}
 H &=& -t\sum_{\langle i,j \rangle, \sigma} 
      ( b_{i}b^{+}_{j}f^{+}_{i,\sigma}f_{j,\sigma}
       +b_{j}b^{+}_{i}f^{+}_{j,\sigma}f_{i,\sigma} )                           
                                             \nonumber  \\
     &-& \frac{J}{2}\sum_{\langle i,j \rangle}
       b_{i}b^{+}_{i}b_{j}b^{+}_{j}
       ( f^{+}_{i,\downarrow}f^{+}_{j,\uparrow}
        -f^{+}_{i,\uparrow}  f^{+}_{j,\downarrow} )
       ( f_{j,\uparrow}f_{i,\downarrow}
        -f_{j,\downarrow}  f_{i,\uparrow} )
                                             \nonumber  \\
     &-& \mu_{0}\sum_{i, \sigma}b_{i}b^{+}_{i}f^{+}_{i,\sigma}f_{i,\sigma}
     +\sum_{i}\lambda_{i}(b^{+}_{i}b_{i}+\sum_{\sigma}
                       f^{+}_{i,\sigma}f_{i,\sigma}
               -1)  \; .
                                                           \label{eq3}
\end{eqnarray}
Now the no-double-occupancy constraint $n_{i} \le 1$ becomes
an equality condition in terms of a spinon and a holon,
$b^{+}_{i}b_{i}+\sum_{\sigma}f^{+}_{i,\sigma}f_{i,\sigma}=1$.
The last term is to impose this condition through a Lagrange 
multiplier $\lambda_{i}$.

    In the spirit of a mean-field approximation, the terms with 
more than two operators should be decoupled in all possible ways.
In principle, we may consider infinitely many species of order parameters.
(infinitely) Many of them are irrelevant, namely, they do not have 
a stable mean-field solution, while
(infinitely many) the others are relevant.  
In this situation, guidance from our physical intuition 
or more likely from experiments
is extremely helpful to find most important leading correlations.
The phase diagram found in high temperature superconductors  
has given us a clear 
answer\cite{Comment10} to this question:
AF and $d$-wave pairing correlations.
Thus, in this paper we consider two order parameters
with broken symmetry,
$m$ and $s$ for AF and $d$-wave orders, respectively.
A similar decoupling was previously considered 
by other groups.\cite{Inui:1988,Inaba:1996}
We also consider fermionic and bosonic exchange couplings, 
$\langle f^{+}_{i,\sigma}f_{j,\sigma} \rangle$ and 
$\langle b^{+}_{i}b_{j} \rangle$, respectively.
In order to simplify calculations as well as to avoid any possible double 
counting problem, we do not give a dynamical aspect to holons.
The presence of holons, which are introduced to keep track of empty sites,
is taken into account as enforcing correlated hopping of spinons
in the hopping term.
In a mean-field approximation, 
$\langle b^{+}_{i}b_{j} \rangle$ is approximated as 
$x=1-n$ and 
$\lambda_{i}$ is replaced by $\lambda$.
Now the three order parameters and the holon hopping amplitude are defined as 
\begin{eqnarray}
   m &=& \frac{1}{2}(-1)^{i}
         \langle f^{+}_{i,\uparrow}
                 f    _{i,\uparrow}
                -f^{+}_{i,\downarrow}
                 f    _{i,\downarrow} \rangle   
                                             \nonumber  \\
   &=& \frac{1}{2N}\sum_{\vec{k}}
         \langle f^{+}_{\vec{k}+\vec{Q},\uparrow}
                 f    _{\vec{k},\uparrow}
                -f^{+}_{\vec{k}+\vec{Q},\downarrow}
                 f    _{\vec{k},\downarrow} \rangle   \; ,
                                             \nonumber  \\
   s &=& \frac{1}{N}\sum_{\vec{k}}\phi_{d}(\vec{k})
         \langle f_{\vec{k},\downarrow}f_{-\vec{k},\uparrow} \rangle  \; ,
                                             \nonumber  \\
   \Delta_{f} &=& \langle f^{+}_{i,\sigma}f_{j,\sigma} \rangle
            = \frac{1}{2N}\sum_{\vec{k}}\phi_{s}(\vec{k})
        \langle f^{+}_{\vec{k},\sigma}f_{\vec{k},\sigma} \rangle  \; ,
                                             \nonumber  \\
   \Delta_{b} &=& \langle b^{+}_{i}b_{j} \rangle
            \simeq x=1-n  \; ,
                                                           \label{eq4}
\end{eqnarray}
where 
\begin{eqnarray}
  \phi_{d}(\vec{k}) &=& \cos k_{x}-\cos k_{y}   \; ,
                                             \nonumber  \\
  \phi_{s}(\vec{k}) &=& \cos k_{x}+\cos k_{y}  \; .
                                                           \label{eq5}
\end{eqnarray}
$\vec{Q}$ is the AF wave vector $(\pi,\pi)$ in two dimensions and 
$N$ the total number of lattice sites.

   In terms of the above parameters,
the mean-field Hamiltonian is written
\begin{eqnarray}
 H_{\mbox{MF}} &=& \sum_{\vec{k}, \sigma}\varepsilon(\vec{k}) 
                  f^{+}_{\vec{k},\sigma} 
                  f_{\vec{k},\sigma}
                                             \nonumber  \\
        & &  -2Jm\sum_{\vec{k}}(f^{+}_{\vec{k}+\vec{Q},\uparrow}
                              f    _{\vec{k},\uparrow}
                             -f^{+}_{\vec{k}+\vec{Q},\downarrow}
                              f    _{\vec{k},\downarrow})
                                             \nonumber  \\
        & &  -Js\sum_{\vec{k}}\phi_{d}(\vec{k})( 
              f^{+}_{\vec{k},\uparrow}  f^{+}_{-\vec{k},\downarrow}
             +f    _{\vec{k},\downarrow}f_{-\vec{k},\uparrow})
           +F_{0} \; 
                                                           \label{eq6}
\end{eqnarray}
where 
\begin{eqnarray}
  \varepsilon(\vec{k}) &=& -\phi_{s}(\vec{k})(J\Delta_{f}+2t\Delta_{b})-\mu
                                                          \; ,
                                             \nonumber  \\
   F_{0} &=& N(2Jm^{2}+Js^{2}+8t\Delta_{b}\Delta_{f}+2J\Delta^{2}_{f}-\mu)
                                                          \; ,
                                             \nonumber  \\
   \mu &=& Jn+\mu_{0}-\lambda  \; .
                                                           \label{eq7}
\end{eqnarray}
By introducing  
a four component field operator 
$\Psi^{+}_{\vec{k}}$
\begin{eqnarray}
\Psi^{+}_{\vec{k}}=(f^{+}_{\vec{k},\uparrow},
                    f_{-\vec{k},\downarrow},
                    f^{+}_{\vec{k}+\vec{Q},\uparrow},
                    f_{-\vec{k}-\vec{Q},\downarrow})  \; ,
                                                           \label{eq8}
\end{eqnarray}
Eq. ~(\ref{eq6}) may be written in a more compact form 
\begin{eqnarray}
     H_{\mbox{MF}} = \sum^{'}_{\vec{k}}\Psi^{+}_{\vec{k}}
                     M_{\vec{k}}\Psi_{\vec{k}}
       +F_{0}  \; .
                                                           \label{eq9}
\end{eqnarray}
The prime symbol on the summation requires the summation of wave vectors
in {\em half} of the first Brillouin zone,
in order to take into account the doubling of a magnetic unit cell
in the presence of (commensurate) AF order.
The matrix $M_{\vec{k}}$ is given as
\begin{eqnarray}
M_{\vec{k}}= \left(  \begin{array}{cccc}
 \varepsilon(\vec{k}) & -Js\phi_{d}(\vec{k}) & -2Jm & 0 \\
 -Js\phi_{d}(\vec{k}) & -\varepsilon(\vec{k}) & 0 & -2Jm \\
 -2Jm & 0 & \varepsilon(\vec{k}+\vec{Q}) & Js\phi_{d}(\vec{k}) \\
  0  & -2Jm & Js\phi_{d}(\vec{k}) & -\varepsilon(\vec{k}+\vec{Q})
           \end{array}
   \right)  \; . 
                                             \nonumber  \\
                                                           \label{eq10}
\end{eqnarray}
The energy eigenvalues of $M_{\vec{k}}$ yield
four energy dispersions $\pm E_{\pm}(\vec{k})$,
\begin{eqnarray}
E_{\pm}(\vec{k})  &=&  [ (\varepsilon^2_{\vec{k}}
        +\varepsilon^2_{\vec{k}+\vec{Q}})/2
        +(2Jm)^2+(Js\phi_{d}(\vec{k}))^2
                                             \nonumber  \\
      & & \pm g(\vec{k}) \;\;
             ]^{1/2} \; ,
                                                           \label{eq11}
\end{eqnarray}
where $g(\vec{k})$ is given as
\begin{eqnarray}
g(\vec{k})=[ (\varepsilon^2_{\vec{k}}-\varepsilon^2_{\vec{k}+\vec{Q}})^2/4
   +((\varepsilon_{\vec{k}}+\varepsilon_{\vec{k}+\vec{Q}})(2Jm)
       ]^{1/2} \; .
                                             \nonumber  \\
                                                           \label{eq12}
\end{eqnarray}

   The free energy is easily obtained either from the trace formula
or from the Feynman theorem
\begin{eqnarray}
F=-2T\sum^{'}_{\vec{k}}\sum_{\alpha=\pm}\log
               (2\cosh \frac{E_{\alpha}(\vec{k})}{2T})+F_{o}  \; .
                                                           \label{eq13}
\end{eqnarray}
Now three mean-field equations are obtained by the stationary
condition of $F$ with respect to the corresponding order parameters,
$\frac{\partial F}{\partial m}
=\frac{\partial F}{\partial s}
=\frac{\partial F}{\partial \Delta_{f}}=0$, and
one more unknown constant $\mu$ is determined by the thermodynamic
relation $n=1-x=-\frac{\partial F}{\partial \mu}$.
The resulting four equations are
\begin{eqnarray}
& m & =  \frac{1}{2N}\sum^{'}_{\vec{k}}\sum_{\alpha=\pm}
     \Bigl\{
           (2Jm)
           + \alpha\frac
 {(\varepsilon_{\vec{k}}+\varepsilon_{\vec{k}+\vec{Q}})^{2}(2Jm)}{2g(\vec{k})}
     \Bigr\}
                                             \nonumber  \\
  & &\times \frac{1}{E_{\alpha}(\vec{k})}
           \tanh(\frac{\beta E_{\alpha}(\vec{k})}{2})  \; ,
                                             \nonumber  \\
                                              \\
                                                           \label{eq14}
& s & =  \frac{1}{2N}\sum^{'}_{\vec{k}}\sum_{\alpha=\pm}\phi^{2}_{d}(\vec{k})
           (Js)
 \frac{1}{E_{\alpha}(\vec{k})}\tanh(\frac{\beta E_{\alpha}(\vec{k})}{2})  \; ,
                                             \nonumber  \\
                                               \\
                                                           \label{eq15}
& \Delta_{f} & =  \frac{1}{4N}\sum^{'}_{\vec{k}}
                  \sum_{\alpha=\pm}\phi^{2}_{s}(\vec{k})
           (J\Delta_{f}+2t\Delta_{b})
     \Bigl\{
           1+\alpha \frac{2\mu^{2}}{g}
     \Bigr\}
                                             \nonumber  \\
  & & \times  \frac{1}{E_{\alpha}(\vec{k})}
              \tanh(\frac{\beta E_{\alpha}(\vec{k})}{2})  \; ,
                                             \nonumber  \\
                                              \\
                                                           \label{eq16}
& n & =  1-\frac{1}{2N}\sum^{'}_{\vec{k}}\sum_{\alpha=\pm}
     \Bigl\{
           (\varepsilon_{\vec{k}}+\varepsilon_{\vec{k}+\vec{Q}})
                                             \nonumber  \\
     & &    + \alpha\frac
           {
           (\varepsilon_{\vec{k}}+\varepsilon_{\vec{k}+\vec{Q}})
 (\varepsilon_{\vec{k}}-\varepsilon_{\vec{k}+\vec{Q}})^2}{2g(\vec{k})}
                                             \nonumber  \\
     & &    + \alpha\frac
           {2(2Jm)^{2}
           (\varepsilon_{\vec{k}}+\varepsilon_{\vec{k}+\vec{Q}})}
           {g(\vec{k})}
     \Bigr\}
     \frac{1}{E_{\alpha}(\vec{k})}\tanh(\frac{\beta E_{\alpha}(\vec{k})}{2})
                                \; .
                                             \nonumber  \\
                                                           \label{eq17}
\end{eqnarray}

   Before presenting our results, several comments are in order concerning 
the mean-field Hamiltonian and equations.
First, $b_{i}b^{+}_{i}b_{j}b^{+}_{j}$ in the second term and    
$b_{i}b^{+}_{i}$ in the third term of Eq. ~(\ref{eq3}) are replaced by 
unity. Extra decoupling of these factors may double count what spinons
have already taken care of.  
Anyhow replacing those factors by unity is expected to 
be a reasonable approximation, as long as doping $x=1-n$ is not high.
Second, in this paper a uniform bond order of $\Delta_{f}$ is 
assumed, namely, 
$\langle f^{+}_{i,\sigma}f_{j,\sigma} \rangle$ is independent of 
a relative direction of a bond $\langle i,j \rangle$.
$\pi$-flux and staggered flux phases   
whose $\langle f^{+}_{i,\sigma}f_{j,\sigma} \rangle$ depends 
on a relative direction of a bond,
were found to be stable only  
at very small doping for $t/J > 1$.\cite{Ubbens:1992}
Third, the mean-field decoupling
$ \langle f_{i,\downarrow}
           f_{j,\uparrow}
          -f_{i,\uparrow}
           f_{j,\downarrow} \rangle $
in general induces superconductivity
with extended $s$-wave symmetry as well as with $d$-wave symmetry.
As noted by Inui {\it et al.}\cite{Inui:1988},
the former is strongly suppressed in the presence of a finite staggered 
magnetization.
As will be shown shortly, 
mean-field AF order is also found to exist in a large 
phase space near half-filling. This may justify neglecting 
superconductivity with extended $s$-wave symmetry in the important region 
of the phase space.
Last, in this paper spin-triplet order parameter 
$\langle f_{\vec{k}+\vec{Q},\uparrow}
f_{-\vec{k},\downarrow} \rangle$ is not explicitly considered. 
In a mean-field decoupling scheme, there is no such term which directly 
induces spin-triplet order parameter.
However, it is dynamically generated when two mean-field orders 
$m$ and $s$ coexist, without affecting $m$ and $s$.\cite{Kyung:2000-1}
Although spin-triplet order parameter is not explicitly mentioned here, 
it appears in the coexistence region of mean-field AF and SC phases.
\section{Results and discussions}
\label{section3}

\begin{figure}
 \vbox to 7.0cm {\vss\hbox to -5.0cm
 {\hss\
       {\includegraphics{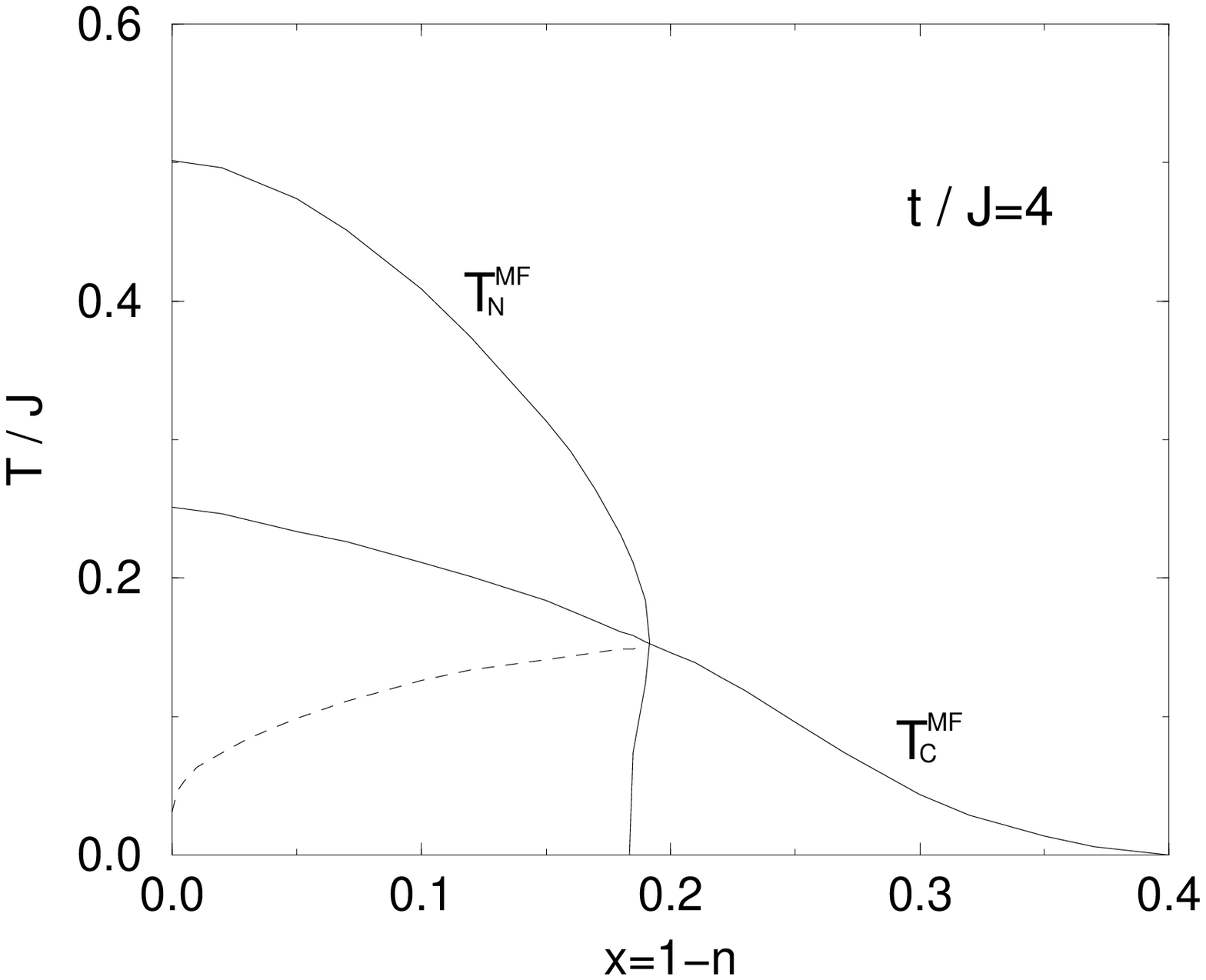}
       }
  \hss}
 }
\caption{Calculated phase diagram in doping ($x=1-n$)
         and temperature ($T$) plane for $t/J=4$. 
         $T^{\mbox{MF}}_{N}$ and $T^{\mbox{MF}}_{c}$ are mean-field AF and SC 
         ordering temperatures, respectively.}
\label{fig1}
\end{figure}

   In Fig. ~\ref{fig1} our calculated mean-field phase diagram is 
presented for $t/J=4$.
Its variation to $t/J$ (=3-4) is negligible. 
Near half-filling mean-field AF and SC orders coexist, while 
far away from half-filling only the latter prevails.
The overall result is qualitatively similar 
to what other groups\cite{Inui:1988,Inaba:1996} obtained before.
It is also similar to our previous result\cite{Kyung:2000-1} based 
on a phenomenological model in which SC order and AF order come  
from different interaction terms.
At this point, we should point out reasons for rejecting 
the unphysical result (dashed curve) of the calculated 
mean-field SC order near half-filling.  
Other groups\cite{Inui:1988,Inaba:1996} also obtained a similar result for
the mean-field $T_{c}$ near half-filling.
Due to the following reasons (both theoretical and experimental), 
we take the solid curve 
as more appropriate mean-field $T_{c}$, which is obtained by setting 
$m$ to zero.
In this respect, the solid curve may be the upper bound  
of the correct mean-field $T_{c}$.

   Theoretically there are two reasons why the reducing of SC order parameter
near half-filling is unphysical.
First, as also noted by Inui {\it et al.}\cite{Inui:1988},
in a (uniform) mean-field approximation  
holes have direct overlap with a staggered AF 
order and suffer from strong time reversal symmetry breaking. 
It causes a rapid destruction of the mean-field SC order near half-filling.
However,
the local deformation of 
AF order in the immediate vicinity of holes, which is 
absent in a typical mean-field approximation,
enables the holes to avoid direct overlap with the AF order.
Consequently mean-field SC and AF orders can coexist without sacrificing 
the energy gain through the local deformation of AF order or 
through a microscopic separation of the two orders like in  
stripes.\cite{Zaanen:1989,Tranquada:1997}

   Second, the unphysical result comes from the limitation of the 
$t-J$ model near half-filling.
At half-filling, the kinetic energy term of the $t-J$ model collapses.
In fact this is directly responsible for the degeneracy of 
superconductivity with $d$-wave symmetry and extended $s$-wave 
symmetry at half-filling, as first noticed by Kotliar.\cite{Kotliar:1988}
However, the collapsing of the kinetic energy term
does not happen in the Hubbard model at any filling,
as long as $t/U$ is kept finite.
As noted in our previous work,\cite{Kyung:2000-2}
the energy dispersion of a hole is given by
\begin{eqnarray}
  \sqrt{(2t)^{2}(\phi_{s}(\vec{k}))^{2} + (U/2)^{2}
        + \Delta^{2}(\phi_{d}(\vec{k}))^{2}}
                                             \nonumber  \\
                                                           \label{eq18}
\end{eqnarray}
for a half-filled Hubbard band with $ U > W=8t $.
$\Delta \phi_{d}(\vec{k})$ is the $d$-wave mean-field SC gap.
For a realistic strength of $U$ ($U \sim 1.5W$),
$\Delta$ was found to be $2.07t$.
Even for $U \gg t$ it saturates to be $2.69t$.
Thus the characteristic energy scale for the hopping term, $2t$, 
is never much smaller than $\Delta$.
As a result, 
the degeneracy of
superconductivity with $d$-wave symmetry and extended $s$-wave
symmetry at half-filling does not happen in the Hubbard model. 
This limitation of the $t-J$ model near half-filling
is ascribed to an inconsistent treatment of the hopping 
and the superexchange terms in the $t-J$ model.
The former is obtained in the $U=\infty$ limit, while 
the latter in the finite $U=4t^{2}/J$ limit.
When the same limit is applied to the Hubbard model,
the hopping term in Eq. ~(\ref{eq18}) vanishes when divided by $U$, 
but the last term survives in the finite $U$ limit.
This leads to the same result as the  
$t-J$ model at half-filling.
The consequence of this excessive reducing of the kinetic term of the 
$t-J$ model near half-filling is to enhance localized AF correlations.
It causes SC order to lose in competition with the AF order, 
even after the local deformation of AF order is properly taken into  
account.

   One strong evidence supporting this argument is provided by failure in
capturing pairing correlations in the exact diagonalization (ED)
study of the $t-J$ model at half-filling.\cite{Leung:1997}
In spite of its exact and unbiased nature, the obtained energy
dispersion is similar to what would be found only by AF
correlations.
This is contrasted with the experimentally observed
energy dispersion in insulating cuprates,\cite{Wells:1995,Ronning:1998}
namely, a $d$-wave-like modulation of the insulating gap.
As a result, a reasonable agreement with the experimental result
is achieved only by introducing unjustified fitting parameters
such as the $t'$ and $t"$ terms into the $t-J$ model.
In fact these terms act like enhancing the kinetic energy term.
Yet another evidence comes from the U(1) gauge theory of the $t-J$ model
by Ubbens and Lee.\cite{Ubbens:1994}
These authors found that the spin-gap (or pseudogap) phase is
completely destroyed by gauge-field fluctuations near half-filling.
In this respect, going to the SU(2) formulation may not help to resolve
the problem.
This limitation of the $t-J$ model near half-filling
may be overcome by using
a fully systematic $t/U$ expansion in the Hamiltonian, the wave function
and all the operators from which the single particle Green's function,
optical conductivity and so on are
defined.\cite{Stephan:2000,Eskes:1994}
If this problem of the $t-J$ model near half-filling
is corrected, we believe that the exotic phases 
found near half-filling do not exist.

   From experimental point of view, two recent 
ARPES experiments\cite{Wells:1995,Ronning:1998}
dictate the presence of strong pairing correlations at half-filling.
An ARPES experiment for an insulating cuprate
Sr$_{2}$CuO$_{2}$Cl$_{2}$\cite{Wells:1995}
clearly shows that
the near isotropy and the overall band dispersion
along $(\pi/2,\pi/2)-(\pi,0)$ and
$(\pi/2,\pi/2)-(0,0)$ cannot be explained by considering
only AF order or its fluctuations.
Furthermore
a $d$-wave-like modulation of the insulating gap
in Ca$_{2}$CuO$_{2}$Cl$_{2}$\cite{Ronning:1998}
is totally mysterious from that point of view.
Furthermore numerous experiments\cite{Timusk:1999} have shown that 
the pseudogap temperature $T^{*}$ increases with decreasing doping 
all the way down to half-filling.
It is believed that mean-field $T_{c}$ has a similar doping dependence with  
$T^{*}$.

   Now we are in a position to point out the weaknesses of a mean-field 
approximation in low dimensions.
Based on this observation and some exact results in the Hubbard model,
we will draw useful information from the slave boson mean-field theory 
of the $t-J$ model.
First of all it is of great importance to note that 
in a mean-field approximation long-range order
already sets in, when
the corresponding correlation length reaches roughly one
lattice spacing.
This forces the above mean-field phase line to be interpreted as
the onset 
of the corresponding short-range correlations.
The question that mean-field order can become  
truly long-range order or not, depends 
on whether the correlation length diverges or not with lowering 
temperature.
In this respect,
a potential location of AF long-range order is only at half-filling
where the AF correlation length logarithmically diverges at low 
temperature.
Away from half-filling, the AF correlation length initially grows 
below $T^{\mbox{MF}}_{N}$ and then saturates. 
This tells that the phase diagram of mean-field AF order can be viewed as
the presence of short-range AF correlations 
for $x \leq x_{c} \simeq 0.18-0.19$
at low temperature.
If there were AF long-range order in the model,
it would be at half-filling and at zero temperature
due to the Mermin-Wagner theorem.\cite{Mermin:1966} 

   Since the paring correlation length logarithmically diverges 
{\it at any filling} for $x \simeq 0.35$ with decreasing temperature, 
in principle SC long-range order may appear from half-filling all the 
way up to $x \simeq 0.35$.
In this paper, however, we argue that {\it SC long-range order does not
occur at any filling strictly in two dimensions.}
Below $x=x_{c}$ where the short-range AF 
correlations are present,
the AF fluctuations create {\em locally} the spin density wave (SDW) state.
It causes the breaking of time-reversal symmetry and thus of SC 
long-range phase coherence.
Above $x=x_{c}$ where the paring correlations 
are relatively weak,\cite{Comment20}
quantum fluctuations can easily destroy the SC long-range order.
This interpretation is consistent with 
the rigorous result
by Su and Suzuki.\cite{Su:1998}. 
These authors proved 
the nonexistence of
d$_{x^{2}-y^{2}}$ superconductivity (long-range order)
at any nonzero temperature in
the two-dimensional Hubbard model.
Note that the proof by Su and Suzuki does not exclude increasing
pairing correlations with decreasing temperature
just like the AF
correlations for a half-filled Hubbard band.
The readers should not be confused with SC long-range order
$\langle \Delta_{g}(0) \rangle \ne 0$ and with pairing correlations
$\langle \Delta^{+}_{g}(0)\Delta_{g}(0) \rangle >
 \langle \Delta^{+}_{g}(0)\Delta_{g}(0) \rangle_{0}$.
$\langle \cdots \rangle_{0}$ means an expectation value evaluated
in the ground state of noninteracting electrons.\cite{Comment35}

   This feature of pairing correlations also 
dictates the mean-field SC phase line
to be interpreted
as the onset ($T^{*}$) of short-range pairing correlations
(pseudogap)\cite{Comment38} instead of as true long-range order.
The pairing correlations extend all the way up to $x \simeq 0.35$.
It makes the crossover region of pairing correlations broad
with respect to $x=x_{c}$ at low temperature.
We further argue that strictly in two dimensions 
SC and AF long-range orders are absent even at
zero temperature, since 
the two correlations may act 
like frustrating the long-range coherence 
of the other correlations.\cite{Comment40}
\section{Universal curve of $T_{c}/T^{\mbox{max}}_{c}$ and reduced 
         superfluid density}
\label{section4}

\begin{figure}
 \vbox to 7.0cm {\vss\hbox to -5.0cm
 {\hss\
       {\includegraphics{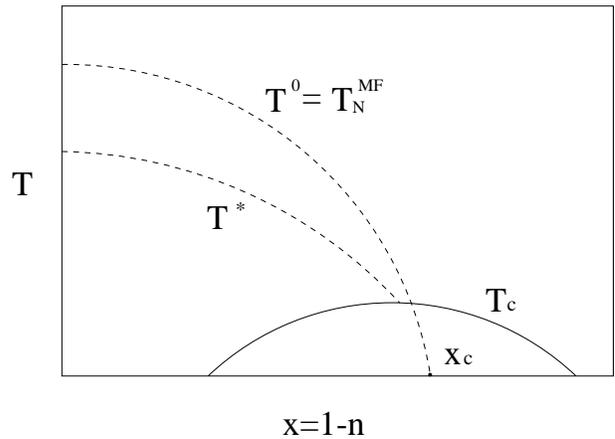}
       }
  \hss}
 }
\caption{Schematic phase diagrams in doping ($x=1-n$)
         and temperature ($T$) plane. 
         $T^{o}=T^{\mbox{MF}}_{N}$ and $T^{*}$ are crossover temperatures 
         of AF and pairing correlations, respectively. $T_{c}$ is 
         a temperature in which SC long-range order sets in.
         $x=x_{c}$ is a doping concentration where 
         $T^{0}=T^{\mbox{MF}}_{N}$ and $T^{*}$ nearly vanish  
         in the absence of interlayer coupling.}
\label{fig2}
\end{figure}

     Therefore, the absence of SC long-range order strictly in 
two dimensions strongly suggests that
true SC long-range
order is driven by a pair hopping process along the c-axis
due to interlayer coupling.
This is consistent with the general trend that
the cuprate compounds with more CuO$_{2}$ planes in a unit cell
show higher $T_{c}$.
In the present paper, the interlayer coupling means not only
the coupling between CuO$_{2}$ planes in a unit cell, but also
between other CuO$_{2}$ planes in different unit cells.
When the interlayer coupling is turned on,
a potential location with the highest $T_{c}$
is near $x=x_{c}$ in which
the phase coherence-breaking AF correlations nearly vanish but
the pairing correlations are still robust.
The resulting phase diagram (Fig. ~\ref{fig2})
will look like one where $T^{*}$ falls from
a high value onto the $T_{c}$ line rather than the other where
$T^{*}$ smoothly merges with $T_{c}$ in the slightly overdoped region, as
recently argued by Tallon and Loram.\cite{Tallon:2000}
In our scenario $T_{c}$ is never part of the $T^{*}$ line.
For different compounds, different strength of interlayer pairing hopping
drives SC long-ranger order
in the {\it same} background of the two-dimensional electron system.
This leads to
a universal relation\cite{Zhang:1993}
$T_{c}/T^{\mbox{max}}_{c} = 1-82.6(x-0.16)^{2}$, when $T_{c}$ is
scaled by $T^{\mbox{max}}_{c}$.
Away from half-filling,
the AF correlations manifest their existence most strongly
in the SC state, because
they easily destroy the SC long-range phase coherence.
Our scenario also predicts that due to scatterings with AF fluctuations,
quasiparticle scattering rate remains finite for $x \leq x_{c}$
even in a clean sample and at $T=0$.
For $x \leq x_{c}$, in our picture,
the Landau quasiparticle with nonvanishing quasiparticle residue
does not exist in the normal state
due to strong scatterings with pairing and AF fluctuations.
But it can be stabilized in the SC state ($T < T_{c}$) where
its coherence is restored through a pair hopping process along the c-axis.

   The crucial point in proper understanding of the SC state is
not the validity of the weak coupling BCS theory, but
more importantly whether Cooper pairs are constructed from
antiferromagnetically correlated electrons or not.
In fact
depending on the doping concentration with respect
to $x=x_{c}$,
the SC state can be qualitatively different.
For $x \leq x_{c}$ the SC state has significant AF correlations, while
for $x > x_{c}$ it has virtually no AF correlations, thus
justifying the conventional
BCS theory based on the noninteracting electrons.
With decreasing doping,
the pairing correlations as well as
the phase coherence-breaking AF correlations increase
in underdoped and optimally doped samples ($x \leq x_{c}$).
As a consequence, with decreasing doping
the SC gap amplitude increases, because it is determined by the 
pseudogap size below $T^{*}$.
On the other hand, superfluid density or $T_{c}$ decreases
due to the increasing phase coherence-breaking AF correlations.
The resulting $(\Delta_{d})_{\mbox{max}}/K_{B}T_{c}$ ratio is strongly
doping dependent, monotonically increasing with decreasing doping
for $x \leq x_{c}$. Above $x_{c}$ (overdoping), however, it is expected
that the ratio approaches more or less 
the BCS mean-field value.\cite{Bardeen:1957}
This feature cannot be understood in the absence of AF correlations
which are allowed in the model.

   The effective strength for the SC long-range
order is also strongly doping dependent and is largest near $x_{c}$.
It decreases below $x_{c}$ due to the increasing phase coherence-breaking
AF correlations and also above $x_{c}$ owing to the decreasing
pairing correlations.
In this respect, it is not surprising to find that
the superfluid density and the SC condensation energy have their
maximum values near $x_{c}$, and decrease below
and above $x_{c}$.\cite{Tallon:2000}
In the present scenario,
the pseudogap is virtually unchanged by an applied
magnetic field, because the characteristic
energy scale for the pseudogap, $\Delta$, is much larger than
the Zeeman energy.
On the other hand,
the SC long-range order is relatively easily destroyed by it
due to its phase coherence-breaking nature.

   In some of magnetic experiments such as NMR,\cite{Reyes:1991}
another crossover temperature
$T^{0}$ (larger than $T^{*}$) is often identified,
at which Knight shift shows its maximum.
This feature can be easily understood on the basis of the competing
nature of pairing and AF correlations as well as
of the phase diagram (Fig. ~\ref{fig2}).
For $T^{*} < T < T^{0}$, the two correlations compete, while they
grow with decreasing temperature.
In spin-lattice relaxation rate which picks up strongly
the $\vec{q}=\vec{Q}$ component, for $T^{*} < T < T^{0}$
the contribution from the AF correlations dominates that from
the pairing correlations. 
It makes $1/T_{1}T$ keep increasing until
it starts to decrease at $T^{*}$.
On the other hand,
Knight shift which picks up
the $\vec{q}=0$ component and thus is unaware of the growing AF
correlations, is more strongly influenced by the increasing
pairing correlations.
Consequently Knight shift reaches its maximum at $T^{0}$, and starts to
slowly decrease below it and then rapidly decrease below $T^{*}$.

   It is also worthwhile to comment on the SC condensation energy in the
present picture.
The SC long-range order is stabilized only
through a pair hopping process along the c-axis (due to the interlayer
coupling).
It forces the SC condensation energy to come from the lowering of the
c-axis kinetic energy in the SC state.
This interlayer coupling theory was already proposed by Anderson and
others,\cite{Wheatley:1988} and many features are consistent with c-axis
optical measurements.\cite{Basov:1994}
\section{Conclusion}
\label{section5}

   In summary,
we have studied the $t-J$ Hamiltonian in a mean-field approximation 
by taking into account AF and $d$-wave pairing correlations.
Considering the presence of AF fluctuations,
the weaknesses of a mean-field approximation and the limitation of  
the $t-J$ model near half-filling,
we gave a new interpretation to the 
slave boson mean-field theory of the $t-J$ model.
We argued that due to phase coherence-breaking AF
fluctuations and quantum fluctuations, SC long-range order does not  
appear strictly in two dimensions.
$T_{c}$ resulting from the interlayer pairing hopping
can lead to a universal relation,
when $T_{c}$ is
scaled by $T^{\mbox{max}}_{c}$.
Systematic reduction of superfluid density and increase of   
$(\Delta_{d})_{\mbox{max}}/K_{B}T_{c}$ ratio below and near optimal doping 
have their natural explanation in our picture.
Another crossover temperature $T^{0}$ found in 
some of magnetic experiments such as NMR
is also easily understood in the present framework.
\acknowledgements

   The author would like to thank A. M. Tremblay for helpful
discussions throughout the work. 
The present work was supported by a grant from the Natural Sciences and
Engineering Research Council (NSERC) of Canada and the Fonds pour la
formation de Chercheurs et d'Aide \`a la Recherche (FCAR) of the Qu\'ebec
government.
\newpage

\begin{references}
\bibitem{Comment10}    
In this paper we restrict ourselves to a uniform state.
On general grounds, solutions with an inhomogeneous stripe structure
are possible.
\bibitem{Comment20}    
Since at a large doping level $\langle b^{+}_{i}b_{j} \rangle
\simeq x$ underestimates the hopping between spinons,
the obtained mean-field $T_{c}$ is overestimated.
In a more realistic situation,
the mean-field $T_{c}$ should be more strongly suppressed
near and beyond optimal doping.
\bibitem{Comment35}    
$\Delta_{g}(i)$ is a pair operator 
with an appropriate gap symmetry.
Generally $\Delta_{g}(i)$ is defined as
$\Delta_{g}(i)  =  \frac{1}{2}\sum_{\delta}g(\delta)
       (c_{i+\delta,\uparrow}        c_{i,\downarrow}
       -c_{i+\delta,\downarrow}      c_{i,\uparrow})$,
where $g(\delta)$ is an appropriate
gap structure factor in real space.
\bibitem{Comment40}    
Unlike the SC long-range order, 
the absence of AF long-range order at zero temperature
is not so crucial for our argument.
\end{references}
\end{document}